\documentclass[preprint,showpacs,preprintnumbers,amsmath,amssymb]{revtex4}

\usepackage{graphicx}

\begin{document}

\title{Dynamics of the $7\times 7$ DAS-reconstructed Silicon (111) Surface}
\author{Lei Liu}
\author{C. S. Jayanthi}
\author{Shi-Yu Wu}
\affiliation{Department of Physics, University of Louisville, 
Louisville, KY 40292}

\date{\today}

\begin{abstract}
We studied the dynamics of the reconstructed Si (111) surface using a 
total-energy-vibrational-spectrum approach based on a
non-orthogonal tight-binding Hamiltonian. We first established the size 
of the supercell sufficient to yield a reliable determination of surface 
parameters by the structural optimization. The site density of vibrational 
states (SDOS) for the semi-infinite system (the optimized slab on top of 
the bulk) was then calculated using the method of real space Green's function. 
A decomposition scheme for the SDOS that identifies directly coupled 
vibrations for a given mode was also proposed. Our study has uncovered 
and elucidated, for the first time, all the important surface dynamical 
features associated with the dimers, adatoms, stacking faults, and rest atoms.
\end{abstract}

\pacs{68.35.Ja, 71.15.Nc, 68.35.Bs}
\maketitle

The $7\times 7$ Dimer-Adatom-Stacking-Fault (DAS)-reconstructed (111) surface of 
silicon is one of the most complicated surface systems (Fig. 1). The intricacies of this 
reconstructed surface structure render the study of its surface dynamics challenging as 
well as rewarding. Experimentally, an EELS measurement \cite{Da87} of the clean $7\times 7$ surface 
revealed dynamical features at 25-33 meV and 71 meV. A high resolution HAS study of 
the low energy surface modes showed Rayleigh waves (RW) peaked at 8, 10, and 15 
meV \cite{La98}. Theoretical studies of surface vibrations of the reconstructed clean $7\times 7$ silicon 
(111) surface include an {\it ab initio} cluster calculation  \cite{Da87}, a total energy and vibrational 
spectrum-study based on empirical potentials \cite{Li88}, a molecular dynamics (MD) simulation 
study of the phonon spectrum based on Car-Parrinnello fictitious Lagrangian and a 
tight-binding Hamiltonian \cite{Ki95}, and a first-principles finite-temperature MD study of the surface 
dynamics \cite{St95,St96}. The two MD studies had yielded similar surface dynamical features. 
Since both MD studies employed a supercell of 400 atoms and the convergence criterion 
for the force of 0.1 eV/\AA, the comparison suggests that tight-binding approaches will give 
comparable result as that obtained by first-principles approaches, provided that both 
approaches used the same computational framework. However, with the imposed 
inversion symmetry and atoms in the central two layers in the supercell held at bulk 
positions in the two MD studies, only 151 atoms (adatoms plus atoms in the first three 
layers) were allowed to relax in the structural determination or to move in the dynamical 
simulations. Thus the extent of the relaxation in the surface structural determination and 
the coupling of the surface to the bulk vibrational modes are not adequately taken into 
account in the two MD studies. Furthermore, the MD trajectory was followed for only 
1.2 ps in the first-principles MD study, suggesting that the first-principles MD study may 
not have provided sufficiently accurate description of the low energy modes. Because of 
the complexity of the surface structure, experimentalists need to have guiding input to 
efficiently measure surface dynamical features. Therefore, a reliable and comprehensive 
theoretical investigation of the surface dynamics of the clean $7\times 7$-reconstruted (111) 
surface is urgently needed. 

We carried out a total-energy-vibrational-spectrum study of the clean silicon 
reconstructed (111) surface. Our convergence test showed that the energy optimization 
using a supercell composed of 10 layers and adatoms (494 atoms), with atoms in the 
bottom 4 layers held at bulk positions to mimic the effect of the bulk and periodic boundary 
conditions imposed 
horizontally, is sufficient to ensure a reliable surface structural determination. The energy 
optimization was implemented with a non-orthogonal tight-binding (NOTB) Hamiltonian\cite{Me97}
because of the system size (494 atoms). This Hamiltonian has been found to yield the 
energetics and the reconstruction of the clean Si (001) surface in excellent agreement 
with those obtained by DFT calculations as well as experimental results\cite{Li00}. Our 
simulation has yielded structural parameters for the Si (111) surface comparable to 
previous studies\cite{Ki95,St92} with only one notable exception. We found the distance between 
the adatom and the atom in layer 2 directly underneath it to be about 2.97\AA, compared 
with a result of 2.60{\AA} of Ref. [4] and a result of 2.40{\AA} of Ref. [6,9]. This substantial 
difference can be traced to the fact that previous studies [4,9] relaxed only 
adatoms and atoms in the first {\it three} layers (151 atoms) at a convergence criterion 
for the force of 0.1 eV/{\AA}, with only {\it one} layer of 
atoms held at bulk positions while our optimization relaxes 298 atoms with the force criterion 
at 10$^{-3}$ eV/{\AA} and with four layers held at bulk positions.
Thus our optimization is expected to provide a more accurate determination of the surface structure. 
We merged the optimized structure of 
the 10-layer slab with the bulk structure to form the semi-infinite $7\times 7$   reconstructed 
DAS (111) surface system. The vibrational spectrum of this semi-infinite system was 
calculated layer by layer using the method of real space Green's function\cite{Wu95}. In the 
calculation, the force constant matrix for the adatoms and the atoms in the first 6 layers 
(298 atoms) in the semi-infinite system and the coupling force constant matrix between 
atoms in the first 6 layers and those in layers 7-10 (interfacial layers) were computed 
numerically as second partial derivatives of the total energy of the original 10-layer slab 
about its equilibrium configuration. The force constant matrix for the atoms in layers 7-10
and those between these atoms and atoms in subsequent bulk layers were again 
computed numerically as second partial derivatives of the total energy of a slab 
composed of layers 3-10 in the original slab plus 4 bulk layers. The bulk force constant 
matrix was used as that for the remaining atoms in this semi-infinite system. In this way, 
the dynamical matrix of the semi-infinite system was constructed continuously and 
smoothly from the surface region, to the interfacial region, and to the bulk region. The 
vibrations of all the atoms in the system were therefore treated on equal footing. Our 
study has identified and elucidated, for the first time, all the important dynamical features 
associated with the dimers, adatoms, stacking faults, and rest atoms.

Fig. 2 presents SDOS for certain adatoms 
and certain atoms in the first 4 layers to highlight important surface features and illustrate 
the underlying physics. They are all presented in the same scale, i.e., 
$\rho_{i\alpha}(\omega)=-(2\omega/\pi)Im_{\varepsilon\rightarrow 0}G_{i\alpha,i\alpha}(\omega^2+i\varepsilon)$
with $G(\omega^2)=(\omega^2-\phi)^{-1}$, $\phi$ being the mass-normalized 
force constant matrix, $i$ denoting the site, and $\alpha$ the direction, so as to allow meaningful 
comparisons between them. We have calculated SDOSs for atoms in the 21st layer and 
found them to be very close to each other and in good agreement with the bulk DOS 
fitted to experiment \cite{We77}. Hence, in Fig. 2(a) we use the SDOS for an atom in the 21st 
layer as the SDOS for a typical atom in the bulk. The SDOSs in Fig. 2 are all 
decomposed along 
$x(\langle\bar1\bar12\rangle)$, $y(\langle1\bar10\rangle)$, and $z(\langle111\rangle)$ 
directions. Fig. 2(a) shows 
that the SDOSs along $x$, $y$, and $z$ directions for a bulk atom are practically all the same. 
Hence one can use the extent of deviation of SDOSs along $x$, $y$, and $z$ directions from 
each other as a gauge of the surface effect. To provide useful guidelines for the 
experimental investigation, we present in Figs. 2(b) through 2(i) the SDOSs for atoms 
(see Fig. 1) that play significant roles to surface modes. Figs. 2(b) and 2(c) give the 
SDOSs for the corner adatom (COA or atom L0A) and central adatom (CEA or atom 
L0B) in the unfaulted subcell respectively. The SDOSs for atom L0A are similar to those 
of atom L0B, with only a slight shift in peak frequencies. They both have prominent 
surface features at about 14 (in-plane), 17 (in-plane), 62 (out-of-plane), 63 (out-of-plane) 
and 75 (out-of-plane) meV. We did not show SDOSs for adatoms in the faulted subcell 
because they are very similar to the corresponding SDOSs in the unfaulted subcell. In 
fact, the features of the SDOSs for the corresponding sites in both halves are in general 
similar, with a few notable exceptions to be discussed later. A comparison of Figs. 2(c), 
2(f) (SDOSs for atom L2B in layer 2 directly underneath atom L0B), and 2(h) (SDOSs 
for atom L3B directly underneath  atom L2B) immediately identifies the split-off mode at 
75 meV as originating from the $z$-vibration of the compressed pair of atoms L2B and 
L3B and attributable to the 71 meV-mode observed by EELS, as also noted by previous 
theoretical studies \cite{Ki95,St95,St96}. An examination of Figs. 2(b) and 2(d) (SDOSs for the rest atom 
L1C with a dangling bond in the unfaulted subcell of layer 1) reveals that the mode at 63 
meV actually originates from the rest atom L1C while Figs. 2(b) and 2(g) (SDOSs for 
atom L2E in a dimer in layer 2 adjacent to the COA L0A) show that the mode at 62 meV 
originates from the dimer. The rest atom mode at 63 meV can be traced from its source to 
the COA L0A by examining Figs. 2(d), 3(b) (SDOSs of the atom L2G in layer 2 that is a 
nearest neighbor of the rest atom L1C), 3(a) (SDOSs of the atom L1F in layer 1 that is a 
nearest neighbor of both the COA L0A and the atom L2G), and 1(b). The pronounced 
mode at 63 meV with almost equal $x$- and $y$-polarization originating from the rest atom 
L1C is seen to induce a vibration at its neighboring atom L2G polarized almost in the $xz$ 
plane, which in turn excites its neighboring atom L1F to vibrate almost in $y$-direction, 
yielding eventually an almost $z$-polarized vibration for the COA L0A. From Figs. 2(g), 
3(a), and 2(b), the pronounced mode at 62 meV polarized along the dimer axis ($y$-direction)
of the dimer atom L2E is seen to have excited its neighboring atom L1F to 
vibrate with its polarization dominated by the $x$-vibration, which in turn induces the
$z$-polarized vibration at the COA L0A. Our calculation has also shown that these three 
modes are localized surface modes, with the 62 (dimer) and 75 meV (compressed atom 
pair) modes barely discernable at layer 4. However, the rest atom mode at 63 meV 
extends beyond layer 4 and exhibits different polarization in layer 4 in the faulted subcell 
(Fig. 2(i)) as compared to the unfaulted subcell. This can be understood as a feature 
associated with the stacking fault. For the rest atom L1D in the faulted subcell of layer 1, 
because of the stacking fault, there is an atom directly underneath it in the faulted subcell 
of layer 4 whose SDOSs are shown in Fig. 2(i). The 63 meV mode still shows up as an 
in-plane mode polarized almost equally along the $x$ and $y$-axis, the same as the mode at 
its rest-atom origin (see Fig. 2(e)). However, there is no atom in the unfaulted subcell of 
layer 4 directly underneath the rest atom in the unfaulted half of layer 1. This mode 
shows up in the unfaulted half of layer 4 as a mode polarized close to the $x$-axis. The 
effect of the stacking fault, however, shows up most prominently in the low frequency 
strongly $z$-polarized mode at about 16 meV in the SDOSs of the rest atom L1D in the 
faulted half of layer 1. A comparison of Figs. 2(d) and 2(e) reveals that such a mode is 
not present in the SDOSs of the rest atom L1C in the unfaulted half. To shed light on this 
unusual contrast, we decompose the SDOS in terms of off-diagonal density matrix 
elements ($\rho_{i\alpha,j\beta}$) as follows \cite{Li01}:
\begin{equation}
     \rho_{i\alpha}(\omega)=
     \frac
{     \sum_{\beta(\ne\alpha)}\phi_{i\alpha,i\beta}\rho_{i\beta,i\alpha}+
        \sum_{\beta, j(\ne i)} \phi_{i\alpha,j\beta}\rho_{j\beta,i\alpha} }
        {\omega^2-\phi_{i\alpha,i\alpha}     }.
\end{equation}
Because of the presence of $\phi_{i\alpha,j\beta}$, only directly coupled vibrations will contribute to the 
SDOS. Note that $\rho_{i\alpha,j\beta}$ also determines the phase relation between the vibration at $i\alpha$   
and that at $j\beta$. Fig. 4 gives the SDOSs along the $z$-axis in the vicinity of 16 meV 
(solid line), decomposed according to $R_{ij}\le3.15\AA$ (long line segment),
$3.15\AA< R_{ij}\le4.15\AA$ (short line segment), and $4.15\AA< R_{ij}$ (dot-dashed line), 
for rest atoms in the unfaulted (Fig. 4(a)) 
and faulted (Fig. 4(b)) halves respectively. Figs. 4(b) and 2(i) show that the most 
significant contribution to the 16 meV-mode originated from the rest atom in the faulted 
half comes from its coupled vibration with the atom in layer 4 directly underneath it
(distance = 2.29\AA), with some participation of its three nearest neighbors and the three
nearest neighbors in layer 3 of the atom in layer 4. Our 
results also reveals that this mode is localized in the vicinity of the configuration 
composed of the eight atoms mentioned above and anchored by the out-of-phase $z$ 
vibrations of the rest atom L1D and the layer 4-atom L4D underneath it. Since there is no 
such configuration for the rest atom in the unfaulted half, no such mode can exist for the 
rest atom in the unfaulted half (see Fig. 4(a)). Hence the presence and the absence of the 
16 meV-mode at the rest atom L1D in the faulted and the rest atom L1C in the unfaulted 
half respectively is a direct consequence of the stacking fault. Figs. 2(f) and 2(c) show 
another interesting surface mode at about 57 meV. This mode originates as a $x$-polarized 
vibration at the atom L2B in layer 2 directly underneath the CEA L0B in the unfaulted 
half and emerges as a $z$-polarized vibration of CEA L0B, with no appreciable penetration 
to atoms in layer 3 and beyond. The decomposition of the SDOS at atom L2B where the 
mode originates using Eq. (1) indicates that atoms at a distance $R>3.15\AA$ from atom L2B 
give insignificant contributions to the SDOS at atom L2B. We therefore present in Fig. 5(a) 
the decomposition of the SDOS along the $x$-axis of atom L2B in the vicinity of 57 meV 
in terms of off-diagonal density matrix elements only for those atoms at a distance less 
than 3.15{\AA} from atom L2B. It can be seen that almost the entire contribution to $\rho_{L2Bx}$
comes from its three neighboring atoms in layer 1, with no meaningful contribution from 
either atom L0B (directly above) or atom L3B (directly underneath). The absence of the 
contribution from atom L0B to the SDOS of atom L2B and the presence of the 57 meV-
mode as a $z$-polarized vibration in the SDOS of L0B (Fig. 2(c)) is a clear indication that 
the $z$-vibration of atom L0B is not directly coupled to the $x$-vibration of atom L2B. Thus 
the 57 meV-mode is a mode localized in the first two layers. It originates from the 
coupled vibrations of the $x$-polarized vibration of atom L2B with its three neighbors in 
layer 1 and, in turn, induces the $z$-polarized vibration of CEA L0B. One interesting 
feature of the method of analysis based on Eq. (1) may be exemplified by the 
examination of the $z$-polarized mode at 45 meV of Fig. 2(h). This is apparently a mode 
restricted to the immediate neighborhood of atom L3B as there is no evidence of such a 
mode in the SDOSs of its neighboring atoms, including atom L2B, its nearest neighbor. 
We present in Fig. 5(b) the decomposition of $\rho_{L3Bz}$ in terms of 
$\sum_{\beta, j(\ne L3B)}\phi_{L3Bz,j\beta}\rho_{j\beta,L3Bz}/(\omega^2-\phi_{L3Bz,L3Bz}) $
for atoms at distances $R<3.15\AA$ from atom L3B. Figs. 4 and 5 show that there is a 
common feature among all the decomposition curves, namely, the simultaneous 
appearance of pronounced peaks and valleys at the frequency corresponding to 
$\omega=\sqrt{\phi_{i\alpha,i\alpha}}$, 
the natural frequency of vibration for the atom $i$ along the $\alpha$ direction (see 
Eq. (1)). The combined peaks and valleys usually lead to a complete cancellation at the 
natural vibrational frequency and hence no contribution to the SDOS as the atom is not 
isolated. A comparison of Figs. 5(a) and 5(b) however reveals that the natural vibrational 
frequency of atom L3B along the $z$-direction is about 47 meV while that of atom L2B 
along the x-direction is about 45 meV. Thus the $z$-polarized mode at 45 meV of atom 
L3B is the consequence of the perturbation of the natural vibration of atom L3B along 
$z$-direction by the natural vibration of atom L2B along the $x$-direction (see Fig. 5(b)). 
Finally, by comparing SDOSs of corresponding atoms layer-by-layer, we have identified 
RWs at 15, 9, and 7 meV, in good agreement with the experimental result of Ref. [2].

	Our comprehensive study of the dynamics of the $7\times 7$ DAS-reconstructed (111) 
has succinctly explained all the important dynamical features associated with the 
structural features of the reconstructed surface. These findings should provide useful 
guidelines for experimental studies of the dynamics of the $7\times 7$ reconstructed surface.

\begin{acknowledgments}
This work was supported by the NSF (DMR-011284)
and the DOE (DE-FG02-00ER45832) grants. 
\end{acknowledgments}

\vskip 0.2 in
\noindent{\bf FIGURE CAPTIONS}
\vskip 0.1 in
\noindent{Fig.1 A schematic presentation of the $7\times 7$ DAS-reconstructed 
Si (111) surface. The sites used in Figs 2 through 5 are labeled by L0A, L1C,
$\cdots$, a combination of the layer marker (L0, adatom; L1, layer 1, $\cdots$) 
and the position marker (A, B, $\cdots$).}
 \vskip 0.1 in
\noindent{Fig. 2: SDOSs of a typical bulk atom (obtained from an atom in the 
21st layer) (a) and of atoms at sites L0A (b), L0B (c), L1C (d), L1D (e), L2B 
(f), L2E (g), L3B (h), and L4D (i) respectively.}
\vskip 0.1 in
\noindent{Fig. 3: SDOSs of atoms at sites L1F (a) and L2G (b) respectively.}
\vskip 0.1 in
\noindent{Fig. 4: The decomposition of SDOSs $\rho_{L1Cz}$ (a) and $\rho_{L1Dz}$ (b) 
(solid line) using Eq. (1) for $R_{ij}<3.15\AA$ (long line segment),
$3.15\AA <R_{ij}<4.15\AA$ (short line segment), 
and $R_{ij}>4.15\AA$ (dot-dashed line) respectively.}
\vskip 0.1 in
\noindent{Fig. 5: The decomposition of SDOSs $\rho_{L2Bx}$  (a) and  $\rho_{L3Bz}$ (b) 
(solid line) using Eq. (1) for  $R_{ij}<3.15\AA$ respectively. 
In (a), long line segment: coupling between L2B and the atom in layer 1 to the left of L2B; 
short line segment: the average of the coupling to the two atoms in later 1 to the right of L2B; 
dot-dashed line: coupling to atom L3B. In (b), long line segment: coupling between atom L3B to 
atom L2B; short line segment: the average of the coupling to the three atoms in layer 4.}

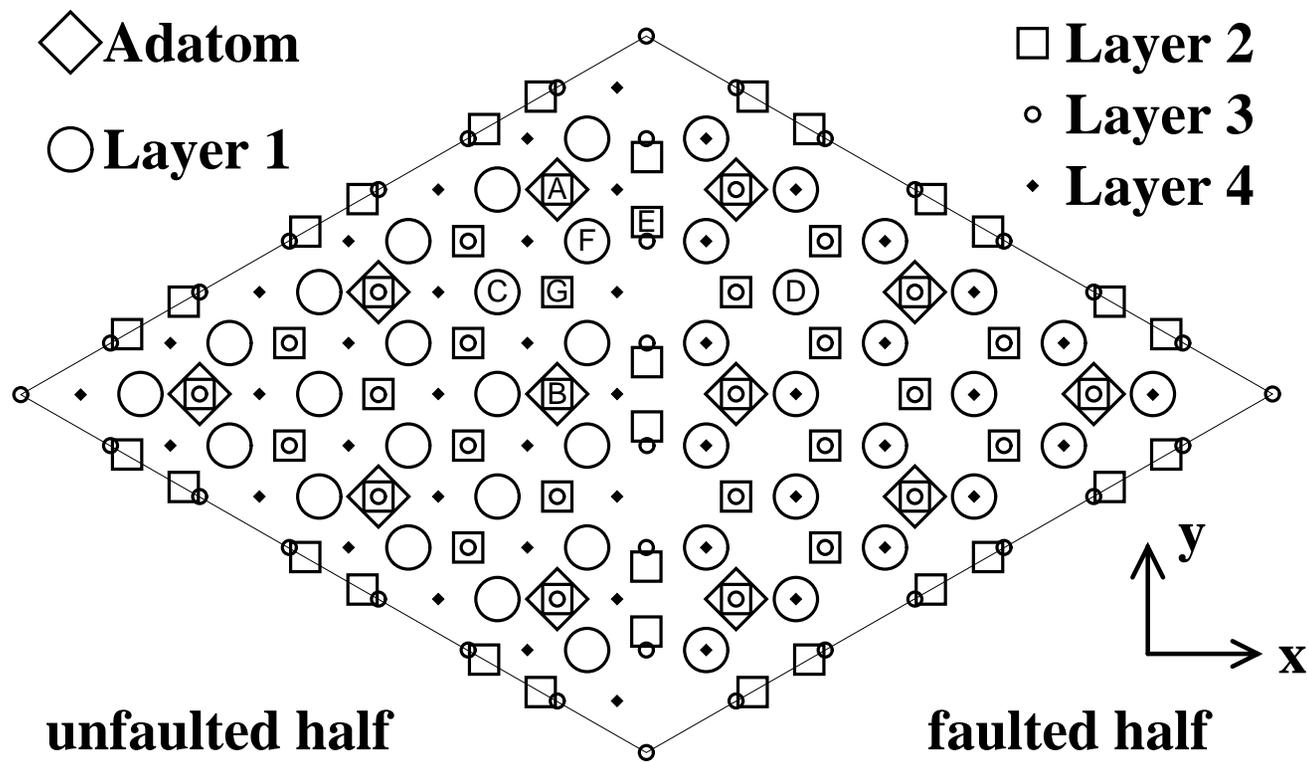
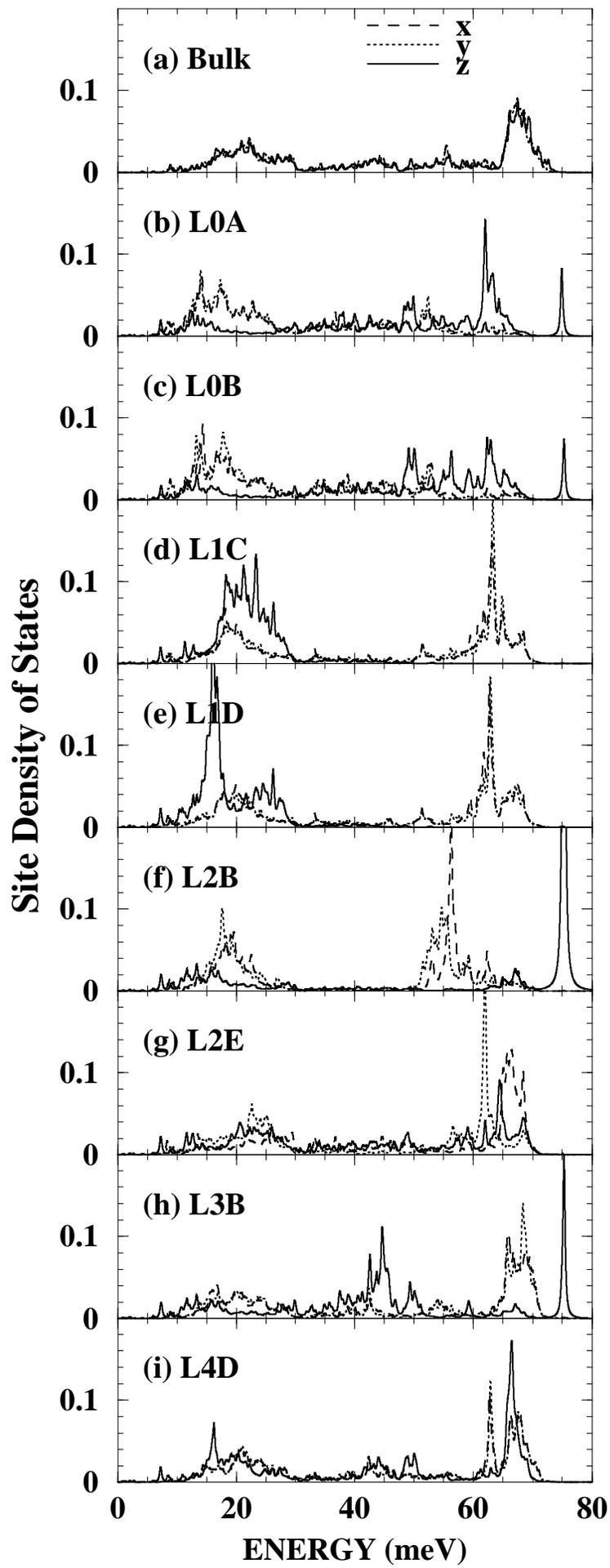
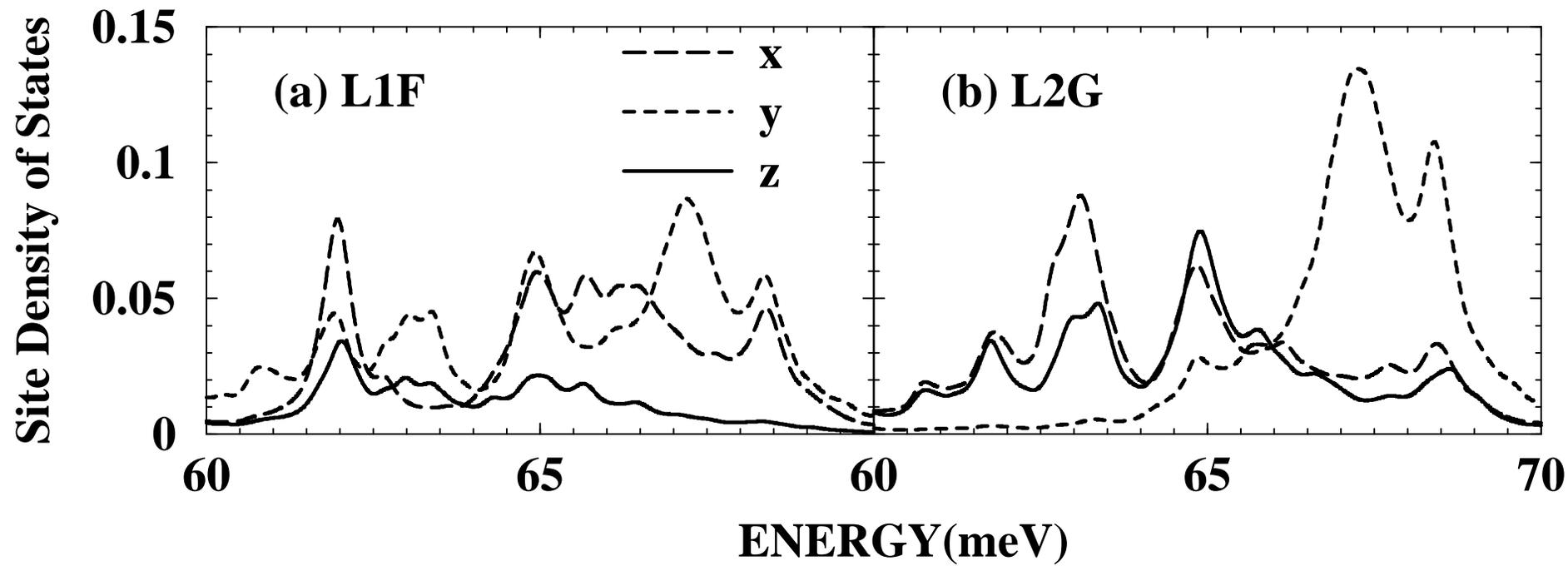
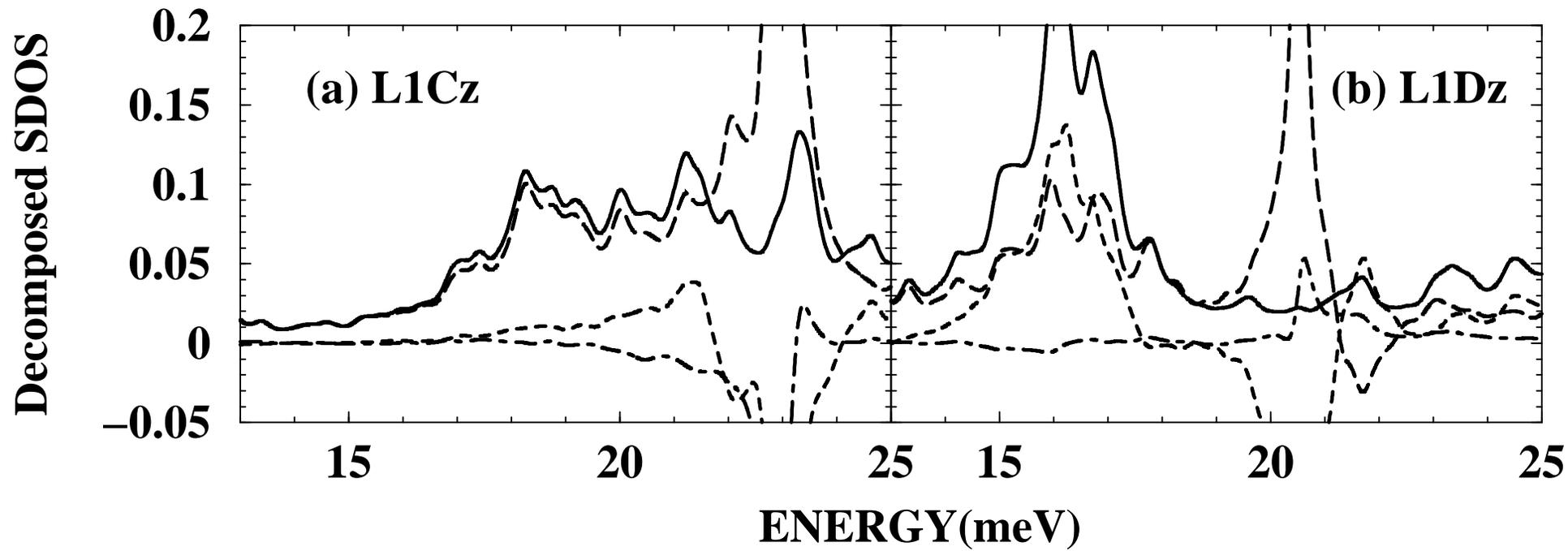
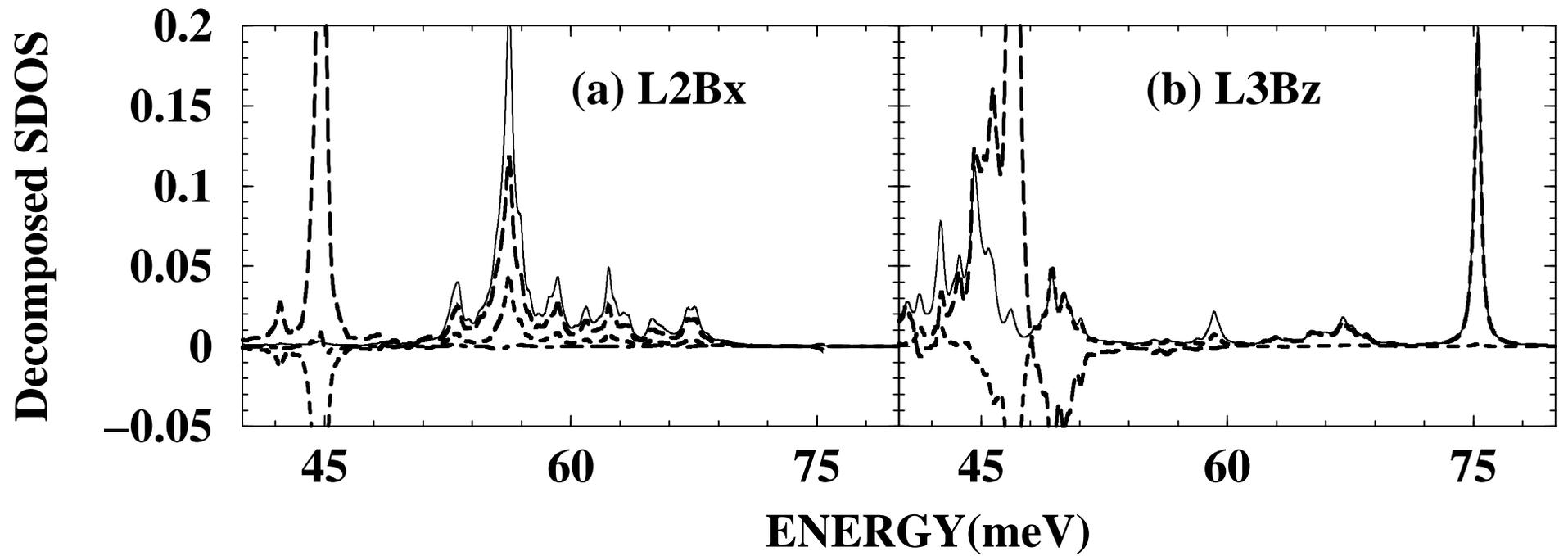
\end{document}